\documentclass[superscriptaddress,twocolumn]{revtex4}
\usepackage{graphicx}
\usepackage{color}

\renewcommand{\vec}[1]{\mbox{\boldmath $#1$}}

\begin{document}

\title{Study of Charge Radii with Neural Networks
} 

\author{Di Wu} 
\affiliation{College of Physics, Sichuan University, Chengdu 610065, China
} 

\author{C.L. Bai}
\affiliation{College of Physics, Sichuan University, Chengdu 610065, China
} 

\author{H. Sagawa}
\affiliation{
RIKEN Nishina Center, Wako 351-0198, Japan}
\affiliation{
Center for Mathematics and Physics,  University of Aizu, 
Aizu-Wakamatsu, Fukushima 965-8560,  Japan}

\author{H.Q. Zhang}
\affiliation{China institute of atomic energy, Beijing, China
}

\begin{abstract}
A feed-forward neural network model is trained to calculate the nuclear charge radii.
The model trained with input data set of proton and neutron number $Z,N$, 
 the  electric quadrupole  transition strength $B(E2)$ from the first excited 2$^+$ state to the ground state, together
with the symmetry energy.  
The model reproduces well not only the isotope dependence of charge radii, but also the kinks of charge radii at 
the neutron magic numbers $N=82$ for Sn and Sm isotopes,  and also $N=126$ for Pb isotopes.
The important role of $B(E2)$ value is pointed out to reproduce the kink of the isotope dependence of charge radii in these
nuclei. Moreover, with the inclusion of the symmetry energy  term in the inputs, the charge radii of Ca isotopes
are well reproduced.  This result suggests  a new  correlation between the symmetry energy and charge radii of Ca isotopes.  
The Skyrme HFB calculation is performed to confirm  the existence of this correlation in a microscopic model.  
\end{abstract}

\maketitle
\section{Introduction}
\label{intro}
Theoretical and experimental studies of the isotopic
changes of ground- and low-lying-state properties 
have  been  studied intensively to elucidate  the evolution of  shell structure,  
 shape coexistence phenomena,  and shape transitions,   over the years
\cite{1a,2a,3a,4a}. In particular, charge radii,  
 and electromagnetic moments
are very sensitive  quantities to extract precise information of  the nuclear structure, 
such as deformation change and the shell evolution along the isotopic or isotonic chains\cite{2a,Thibault,Mueller,Friecke,Kluge,Rossi,Minamisono,Garcia,Gorges,Miller}, 
and to study the neutron skin thickness \cite{ma2016}. 
Different types of
experimental data have been used 
 to obtain  nuclear  radii:  muonic-atom spectra and electron scattering experiments, as well as 
isotope shifts to determine relative radii of neigboring nuclei \cite{5a}. An extensive compilation of the nuclear electromagnetic
moments  can be found  in 
Ref. \cite{6a},   and references therein.
In these researches, the isotope dependence of charge-radii of Ca isotopes shows a particularly unique feature, which raises 
a  challenging quest for the nuclear theory to understand the structure of these isotopes~\cite{Miller,caurier}.  
The behavior of charge radii around the shell closures has been frequently studied in heavy isotopes.  Many isotopes  show  obvious 
kinks at the shell closures, namely,  a sudden change of the slope of charge radii for  the isotopic chain at the magic number.  
In particular, the kink around $N=28$\cite{Garcia,Vermeeren}, $N=82$\cite{Gorges} and $N=126$\cite{Anselment,Cocolios} were extensively  studied.

There are two groups of models for evaluation of charge radius: microscopic and phenomenological ones.  
Shell model and $ab$  $initio$ models provide  reasonable predictions of
the charge radii in light and medium mass nuclei with realistic two-body and three-body interactions\cite{Forssen}.
The energy density functional (EDF)  such as the Hartree-Fock (HF) and HF-BCS (or Bogolyubov) 
models\cite{stoitsov,goriely,nakada06,Reinhard,nakada19} as well as the relativistic mean field models (RMF)\cite{lalazissis,zhao}
 provide  global quantitative descriptions for the charge radii in a wide region of mass table. 
However it is still not quite successful for EDF to fix  optimum  interactions or to choose appropriate functional forms of EDF 
in order to provide precise prediction for the charge radii systematically. 
 Phenomenological models such as the "liquid-drop" model (LDM)\cite{duflo} and Garvey-Kelson relation\cite{GK} 
as well as their developed versions\cite{zhang,wangn,sheng,bao,sun}, in which the isospin dependence, shell effects and 
odd-even staggering are included, are also introduced to 
study the isotope dependence of charge radius.
These phenomenological models work well on the global prediction of the charge-radii,
but it might be difficult for these models to reveal all the important physical quantities in the
fitting,  and to grasp microscopic origins of the model. 

Machine learning  (ML) is one of the most popular algorithms in dealing with complex systems   due to its powerful and convenient inference abilities. 
Neural network, an algorithm of machine learning, has been 
widely used in different fields such as artificial intelligence(AI), medical treatment, and 
physics of complex systems. 
There are many successful applications of machine learning 
in nuclear physics, for examples, 
predictions of the 
nuclear mass~\cite{Utama2016,14,Wu}, charge radii~\cite{ma,utama}, dripline locations~\cite{Neufcourt2018,Neufcourt2019}, 
$\beta$-decay half-lives $T_{1/2}$ ~\cite{15}, the fission product yields~\cite{16}, and the isotopic cross-sections in 
proton induced spallation reactions \cite{ma2020}. 
 Very recently, a multilayer neural network was applied to predict the ground-state and excited  energies with high accuracy~\cite{Lasseri}.  
In the above works, ML was applied to improve   the accuracy of  calculated  results based on the EDF.
In this work, we try to train a description of the nuclear charge radii based directly  on some experimental or quasi-experimental data, 
such as the mass number dependence, shell effects, and deformation.
We are desperately  interesting in finding any other physical quantities which are correlated to the charge radii.
The special attention will be payed to the cases of Ca isotopes. 

To this end, we employ a  standard fully connected feed-forward neural network (FNN),  
which can  build a complex mapping between the input space and output space through multiple compounding of simple non-linear functions.
The framework of the FNN is introduced in section \ref{nn}. The data preprocessing is shown in section \ref{prep}. The results are shown in section 
\ref{results}. Section \ref{summary} is devoted to the summary. 

\section{Neural Network}\label{nn}

\begin{figure}
\centering
\includegraphics[scale=0.26]{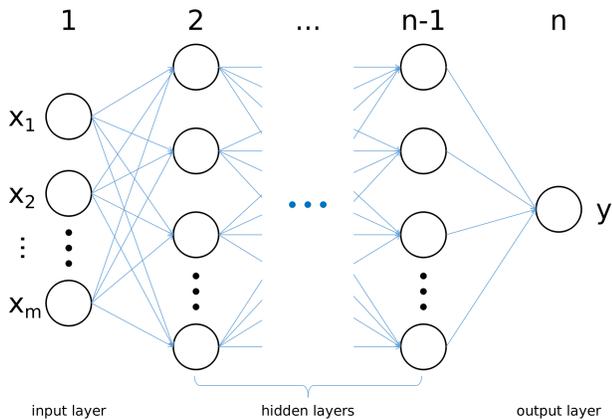}
\caption{(Color online) The framework of FNN, which consists of input layer, hidden layers and output layer. 
The network maps the inputs $\vec{x}_i$ to the corresponding charge radii $y_i=f(\vec{\theta},\vec{x}_i)$ via the weights and biases as well as the activation functions in 
each layer. See text for details.}\label{fnn}
\end{figure}

The framework of FNN is shown in Fig. \ref{fnn}, which is a multilayer neural network consists of input layer, hidden layers 
and output layer.  The structure of the neural network is labelled as $[N_1$, $N_2$, $\ldots$, $N_n]$, where 
$N_i$ stand for neuron numbers of $i^{\mbox{th}}$ layer, and $i=1$ and $n$ represent the input and output layer, respectively. 
In the present work, $N_{1}$=3, and $N_n=1$.
For the hidden layer, the outputs $\vec{h}(\vec{\theta}_{i},\vec{x})$ 
are calculated by a formula 
\begin{equation}
\vec{h}(\vec{\theta}_{i},\vec{x})=\mbox{tanh}(0.01\vec{w}^{(i)}\cdot\vec{h}(\vec{\theta}_{i-1},\vec{x})+0.01\vec{b}^{(i)}),
\end{equation}
which is a $N_i\times 1$ matrix, and the activation functions of hidden 
 layers are taken to be the hyperbolic tangent,  $tanh$.
As the starting point, the outputs $\vec{h}(\vec{\theta}_1,\vec{x})$ of the input layer are 
actually the input data $\vec{x}$.
Finally the correspond output is given by 
\begin{equation}
f(\vec{\theta},\vec{x})=\vec{w}^{(n)}\cdot\vec{h}(\vec{\theta}_{n-1},\vec{x})+\vec{b}^{(n)}, 
\end{equation}
where 
$\vec{\theta}=\{\vec{w}^{(1)},\vec{b}^{(1)},\ldots,\vec{w}^{(n)},\vec{b}^{(n)}\}$
are the network parameters trained by a selected optimizing algorithm.
The number of network parameters is determined by a formula 
\begin{equation}
\begin{array}{l}
N_p=\sum\limits_{i=1}^{n-1}N_i\times N_{i+1}+\sum\limits_{i=2}^{n}N_i. 
\end{array}
\end{equation}

In the training procedure, we use the mean squared error (MSE) as the loss function, 
\begin{equation}
\mbox{Loss}(\vec{y},f(\vec{\theta},\vec{x}))=\frac{1}{N_S}\sum_{i=1}^{N_S}(y_i^{exp}-f(\vec{\theta},\vec{x}_i))^2
\end{equation}
which is used to quantify the difference between model predictions $f$ and experimental values $y^{exp}$.  Here, 
  $N_S$ is the size of  the training set. The learning process is to minimize the 
loss function via a proper optimization method. We use RMSProp  method \cite{18} in this work to 
obtain the optimal parameters $\vec{w}^{(i)}$ and $\vec{b}^{(i)}$,   respectively, for $ i=1,2,\ldots,n$ 
in the network.  
RMSProp is a popular 
alternative to stochastic gradient descent (SGD), which is one of the most widely used training 
algorithms.

In the present work, the deformation and shell effects  on charge radius are 
included by the excitation energy of the first $2^+$ state,  $E_{2_1^+}$.  
We adopt as the data set  all the nuclei for which both the experimental values of $E_{2_1^+}$ and charge radii are available.  
The total data set is 347 nuclei.
Since  the data set is not big enough, we have to choose a  small network structure [3,40,1], which includes 
44 neurons, and involves 201 parameters.

In the training procedure, the training sets are chosen 10 times randomly,  which are equivalent to  10 models for the description
of the charge radii. In this way, we will be able to choose the appropriate model. When training a model with a given training set, the network parameters are initialized randomly
to produce the output. Since the output of the network are related to the initialization, we repeat this process 50 times for each model and average the
output results. Then, the mean value is taken as the result of the 
model. We found that the mean value converges to a certain MSE value after the repeat of 30 times.  

\section{Data preprocessing}\label{prep}

Since the performance of the network  depends tightly on the form of the inputs, the
data preprocessing is essential to arrange the raw inputs.
As it was pointed out in Ref.\cite{angeli2015,cakirli}, there are correlations between the nuclear charge radii and the 
excitation energies of the first excited $2^+$ states,  $E_{2_1^+}$, we choose the energies $E_{2_1^+}$  as a 
raw input.  %
To modulate the irregularity due to the large difference in  the energies $E_{2_1^+}$ between the magic nuclei and its neighboring ones, 
  we smooth the mass number dependence of energies $E_{2_1^+}$  
by the Lorentzian function, 
\begin{equation}\label{Lorentz}
E_{2_1^+}(X_0)=\frac{\Gamma/2}{\pi}\sum_X\frac{E_{2_1^+}(X)}{d_{X_0,X}^2+(\Gamma/2)^2},
\end{equation}
where $X=(N,Z)$,  and the sum of $X$ runs over all adopted nuclei, whose  $E_{2_1^+}$ values are measured. 
Here,  the expansion width $\Gamma$ is set at 7.0 for the best fitting, and $d_{X_0,X}$ has the form
\begin{equation}
d_{X_0,X}=\sqrt{(Z-Z_0)^2+(N-N_0)^2}.
\end{equation}
Because the $B(E2)$ and $E_{2_1^+}$ are  empirically  correlated by the Grodzin's formula 
 \cite{raman},
\begin{equation}  \label{Grodzin}
E_{2_1^+}B(E2)=2.57\times Z^2A^{-2/3},
\end{equation}
we actually take  into account  the calculated $B(E2)$ by Eq. (\ref{Grodzin}) in the ML study.  That means,   the dynamical deformation effect is
included in the present study through $B(E2)$ values.   

In addition, in order to consider the effect of symmetry energy, 
 we introduce a factor $\mathcal{A}(\delta_S)$ which is related to 
the symmetry energy part in the Bethe-Weizs\"{a}cker mass formula\cite{Weizs,bethe},
\begin{equation}\label{af}
\mathcal{A}(\delta_S)=b_1\mbox{exp}(b_2Z^{\beta_1}\delta_S^{\beta_2})+b_3,
\end{equation}
and
\begin{equation}\label{es}
\delta_S=\frac{(N-Z)^2}{A}.
\end{equation}

Furthermore, the $B(E2)$ and $\mathcal{A}(\delta_S)$ are arranged as a input data in the form:
\begin{equation}\label{fb}
g(B(E2),\delta_S)=a_1\mbox{ln}(a_2A^{-1/3}B(E2)\mathcal{A}(Z,\delta_S))\times(a_3Z)^{\alpha}.
\end{equation} 
The optimized values of the parameters   in Eqs.(\ref{af}) and (\ref{fb}) by the trainings are listed in Table \ref{param}. 
\begin{table}
\caption{Values of the parameters in Eqs. (\ref{af}) and (\ref{fb}).} \label{param}
\begin{tabular}{ccccccccc}
\hline
 $a_1$ & $a_2$ & $a_3$ & $b_1$ & $b_2$ & $b_3$ & $\alpha$ & $\beta_1$ & $\beta_2$\\
\hline
 25.0 & 0.1 & 0.05 & 1.5 & $-0.1$ & 1.4 & $-0.5$ & 1.3 & 4.0\\
\hline
\end{tabular}
\end{table}
Thus, the inputs for trainings are $Z$, $N$, and $g(B(E2),\delta_S)$, and the corresponding output is $r_{ch}$. 
 
\section{Results}\label{results}

In the present ML study,  all the charge radii of Ca, Sm, and Pb isotopes
are put   in the testing set in order to check the prediction power of models. In the calculations, the inputs are prepared both with and without the 
symmetry energy input.  That is,  the factor $\mathcal{A}$ is  evaluated by eq.(\ref{af}) or set at  the value 1.0 for all the processes, respectively.
The best models with smallest RMSD values 
 are selected  from each ten models  with and without the symmetry energy input,   after trained with different
training sets. The RMSD values of the two best models are listed in Table.\ref{rmsd}.  We should notice that the RMSD value with the symmetry energy effect
 $\delta_S$ is a lightly larger than that without the $\delta_S$ term.  However we find the substantial improvement of prediction in Ca-isotopes with $\delta_S$ in the testing set as will be seen below.  
\begin{table}
\caption{Minimum RMSD values for the training set and testing set of the models trained with
or without symmetry energy. The values are given in unit of fm.}\label{rmsd}
\begin{tabular}{ccc}
\hline
model & training set & testing set \\
\hline
with $\delta_S$ & 0.0286 & 0.0280 \\
no $\delta_S$     & 0.0266 & 0.0231 \\
\hline
\end{tabular}
\end{table}

\begin{figure}
\centering
\includegraphics[scale=0.7]{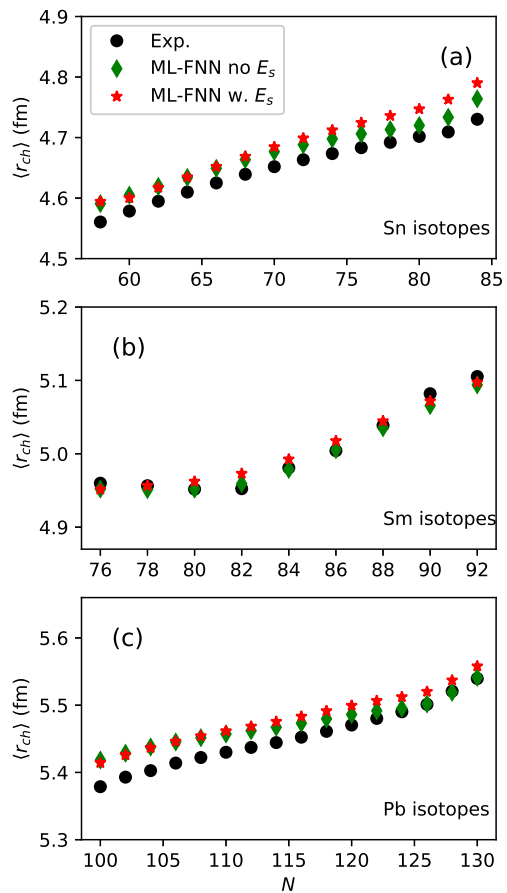}
\caption{(Color online) The results of charge radii of Sn, Sm, and Pb isotopes calculated by the best model trained with and without taking
the symmetry energy  input into account. In the figure, the ML results
obtained with and without the symmetry energy input are labelled by the red stars  and  green diamonds,  respectively.  
The experimental data are taken from Ref. \cite{Gorges,5a,fricke}, which are labelled by the filled black circle. See the text for more details.}\label{Es_comp}
\end{figure}

The results of charge radii of Sn, Sm, and Pb isotopes are shown in the Figure.\ref{Es_comp}. In the figure, the ML results
obtained with and without the symmetry energy input are labelled by the  red stars and green diamonds, respectively. 
The experimental data are taken from Ref. \cite{Gorges,5a,fricke}, which are labelled by the filled circle. Panels (a) and (b) show the charge radii of Sn and Sm
isotopes, respectively.   Both cases with and without the symmetry energy input, the results produce kinks properly at N=82. In panel (c), the charge
radii of Pb isotopic chain are shown.  The kink at N=126 is reproduced by  the two different models. The nuclei
shown in the figure are neutron rich,   and the variation of factor $\delta_S$ is small,  not so much different for each isotope as is given by  Eq. (\ref{es}). 
Accordingly,  the models
trained with or without taking the symmetry energy input  produce the similar results. The reproduction of the kinks at N=82
and 126 indicates that the shell structure and dynamical quadrupole deformation 
are treated properly by the present data processing,  
and the models work well in the heavy nuclei.

\begin{figure}
\centering
\includegraphics[scale=0.7]{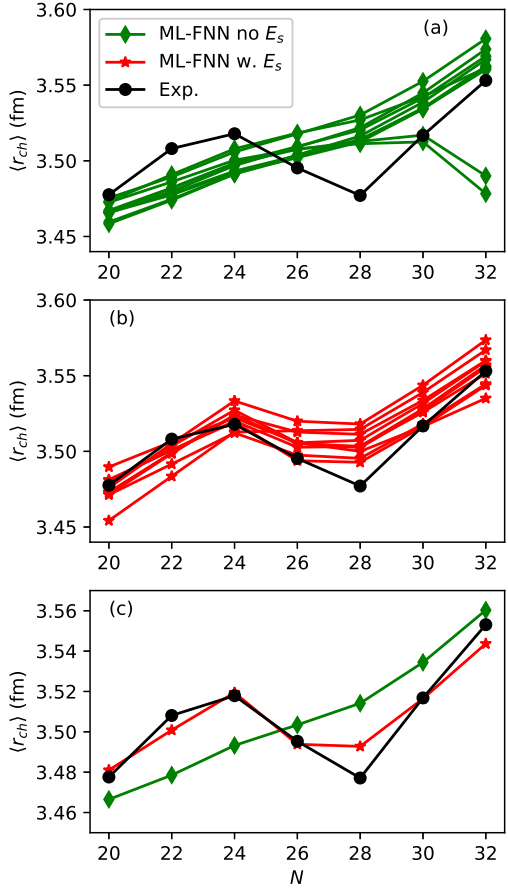}
\caption{(Color online) The same as Fig.\ref{Es_comp}, but for Ca isotopes. (a) The results of the 10 models trained without the symmetry energy.
(b) The results of 10 models trained with the symmetry energy. (c) The best models with the lowest RMSD trained with and without the symmetry energy. 
The experimental data are taken from Ref. \cite{Garcia,5a}. 
See the text for more details.}\label{Ca}
\end{figure}
The charge radii of Ca isotopes are shown in Fig.\ref{Ca}. Experimental data are taken from Ref. \cite{Garcia,5a}, which show a strong kink 
structure at $N=28$, and have a peak between two closed shells at $N=$20 and 28, and then increase rapidly after $N$=28.  
 As shown in the panel (a), the 10 models with random data sets trained without the symmetry energy input 
give  charge radii either linear or parabolic dependences with the increasing of neutron number.   All  results are not   qualitatively consistent
with the experimental data. When the symmetry energy input is  included, as shown in panel (b), the results of 10 models are improved systematically and 
produce not only the kink at $N$=28, but also the peak at $N$=24,   in consistent
with the trend of the experimental data. The best results with the lowest RMSD values for the two kind of models  
with and without symmetry energy input are shown in panel (c).    
This figure indicates that  the symmetry energy input is  critical for the qualitative and quantitative description of the Ca isotopes. 

\begin{figure}
\centering
\includegraphics[scale=0.5]{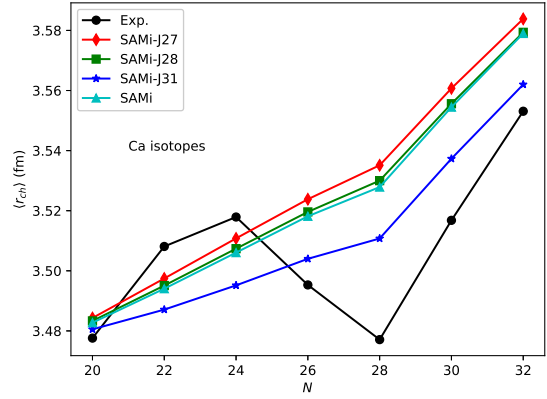}
\caption{(Color online) Charge radii calculated by the Skyrme-HFB with the SAMi-J families and SAMi parameter sets. The diamonds,   the squares,  
the triangles, and  the  stars stand for the results of SAMi-J27, SAMi-J28, SAMi-J31, and SAMi, respectively, while 
the filled circles give the experimental values.}\label{hfb_sami-j}
\end{figure}
To confirm the conjecture derived by the present ML study, we calculate further the correlation between the symmetry 
energy and the charge radii in a microscopic model.
The Hartree-Fock-Bogolyubov (HFB) calculations~\cite{hfbrad}  are performed  with modern Skyrme interactions. 
The energy density functional per nucleon can be expanded up to the second order of the isovector
index $I=(\rho_n-\rho_p)/\rho$ as
\begin{equation}\label{edf}
\varepsilon(\rho_n,\rho_p)=\varepsilon(\rho,I=0)+S(\rho)I^2,
\end{equation}
where $S(\rho)$ is the symmetry energy, which is important for the studies on the properties of finite nuclei and nuclear matter \cite{inakura,colo,li2008,yu2020}. 
The symmetry energy $S(\rho)$ in Eq. (\ref{edf}) is further  expanded around the saturation density as
\begin{equation}
S(\rho)=J(\rho_0)+L(\rho_0)\frac{\rho-\rho_0}{3\rho_0}+K_{sym}(\rho_0)(\frac{\rho-\rho_0}{3\rho_0})^2.
\end{equation}

We calculate the charge radii of Ca isotopes by HFB model 
with SAMi-J family\cite{roca}, which has different behavior for the symmetry energy in the nuclear matter. 
The SAMi-J family was obtained following the fitting protocol of SAMi under 
the constrained conditions;  the incompressibility $K_{\infty}$ and the effective mass $m^*/m$ are fixed 
($K_{\infty}=245$ MeV, $m^*/m=0.675$), and then the symmetry energy at saturation density ($J$) is varied from 27 to 31 MeV 
(SAMi-J27$\sim$SAMi-J31)\cite{roca,roca2012}. 
\begin{table*}
\centering
\caption{Nuclear matter properties of SAMi and SAMi-J family.}\label{sami}
\begin{tabular}{l|ccccc}
\hline
Parameter & $K_{\infty}$ (MeV) & $J$ (MeV) & $L$ (MeV) & $K_{sym}$ (MeV) & $m^*/m$\\
\hline
SAMi & 245.0 & 28.16 & 43.68 & $-119.94$ & 0.675 \\
SAMi-J27 & 245.0 & 27.00 & 30.00 & $-158.04$ & 0.675 \\
SAMi-J28 & 245.0 & 28.00 & 39.74 & $-133.15$ & 0.675 \\
SAMi-J31 & 245.0 & 31.00 & 74.37 & $-37.35$ & 0.675 \\
\hline
\end{tabular}
\end{table*}
Fig. \ref{hfb_sami-j} shows the Skyrme-HFB  results with the SAMi-J family, where $J$=27, 28 and 31 MeV for SAMi-J family, and $J$=28.16 MeV for SAMi. The 
nuclear matter properties of these parameter sets are listed in Table \ref{sami}. 
The figure shows that the greater the  difference of $J$ between two parameter sets is , the more obvious  difference 
in charge radii appears;    the calculated charge radii decrease as the $J$ values increase. Furthermore, 
if the values of $J$ of two parameter sets are similar,   the calculated charge radii are  also close to each other, 
$i.e.$, the charge radii are almost the same for SAMi and SAMi-J28, whose $J$ values are $J=28.16$ and 28 MeV, respectively 
\cite{roca1},  although  the values of slope parameter $L$ are somewhat difference, 
as is shown in Table \ref{sami}. This indicates that the 
symmetry energy term $J$ plays more  important role than the term $L$ in the calculation of charge radii.

\section{Summary}\label{summary}
 A multilayer feed-forward neural network model (ML-FNN)  is applied  to train a model for the discription of the charge radii.
The model is  trained with the input data set of proton number $Z$, neutron number $N$, the $E_{2_1^+}$ 
and the symmetry energy. The model reproduces well not only the slope of isotopic dependence, but also the kink of charge radii at the 
magic numbers $N=82$ and  126. Especially, the inclusion of the symmetry energy input  makes the model better to reproduce qualitatively and quantitatively  
 the charge radii of Ca isotopes. The microscopic Skyrme-HFB calculations show that the symmetry energy term in
the energy density,  particularly,  the $J$ term affects  the isotope dependence of charge radii of Ca isotopes.   The present ML research and the microscopic 
calculation show the new correlation between the symmetry energy and charge-radii in Ca isotopes. Whereas the ML shows that the
symmetry energy input has a remarkable  effect  on the charge radii of Ca isotopes,  
 the HFB calculation also show that the symmetry
energy has  an effect to change the absolute magnitude of charge radii, but the kink structure at $N$=28  is not well reproduced.
The physical implication of present successful ML study is still an open question,  and  needs to be studied in the future.

\begin{acknowledgments}
We would like to thank Prof. Xavi Roca-Maza for  informing  us some details of SAMi-J parameter sets. 
We would also like to thank Prof. Yonghong Zhao for the fruitful discussion of the machine learning.  
This work is supported by the National Natural Science Foundation of China under
Grants No.11575120, and No. 11822504.
This work is also supported by JSPS 
KAKENHI  Grant Number  JP19K03858.  
\end{acknowledgments}


\begin{thebibliography}{99}
\bibitem{1a} B. Cheal and K. T. Flanagan, J. Phys. G: Nucl. Part. Phys. 37,
113101 (2010).
\bibitem{2a}  P. Campbell, I. D. Moore, and M. R. Pearson, Prog. Part. Nucl.
Phys. 86, 127 (2016).
\bibitem{3a} J. L. Wood, K. Heyde, W. Nazarewicz, M. Huyse, and P. van
Duppen, Phys. Rep. 215, 101 (1992).
\bibitem{4a} M. Bender, P.-H. Heenen, and P.-H. Reinhard, Rev. Mod. Phys.
75, 121 (2003).
\bibitem{Thibault}C. Thibault, F. Touchard, S. B\"{u}ttgenbach, R. Klapisch, M.
de Saint Simon, H. T. Duong, P. Jacquinot, P. Juncar, S.
Liberman, P. Pillet, J. Pinard, J. L. Vialle, A. Pesnelle, and
G. Huber, Phys. Rev. C {\bf 23}, 2720 (1981).
\bibitem{Mueller}A. Mueller, F. Buchinger, W. Klempt, E. Otten, R. Neugart,
C. Ekstr\"{o}m, and J. Heinemeier, Nucl. Phys. {\bf A403}, 234 (1983).
\bibitem{Friecke}G. Friecke, C. Bernhardt, K. Heilig, L.A. Schaller, L. Schellenberg, E.B. Shera, C.W.de Jager, At. Data Nucl. Data Tables 60 (1995) 177
\bibitem{Kluge}H.-J. Kluge, Hyperfine Interact. 196 (2010) 295.
\bibitem{Rossi}D. M. Rossi, K. Minamisono, H. B. Asberry, G. Bollen, B. A. Brown, K. Cooper, B. Isherwood, P. F. Mantica, A. Miller, D. J. Morrissey, R. Ringle, J. A. Rodriguez, C. A. Ryder, A. Smith, R. Strum and C. Sumithrarachchi, Phys. Rev. C {\bf 92}, 014305 (2015).
\bibitem{Minamisono}K. Minamisono, D. M. Rossi, R. Beerwerth, S. Fritzsche, D. Garand, A. Klose, Y. Liu, B. Maa{\ss}, P. F. Mantica, A. J. Miller, P. M\"{u}ller, W. Nazarewicz, W. N\"{o}rtersh\"{a}user, E. Olsen, M. R. Pearson, P.-G. Reinhard, E. E. Saperstein, C. Sumithrarachchi and S. V. Tolokonnikov, Phys. Rev. Lett. {\bf 117}, 252501 (2016).
\bibitem{Garcia}R. F. Garcia Ruiz, M. L. Bissell, K. Blaum, A. Ekstr\"omi, N. Fr\"ommgen, G. Hagen, M. Hammen, K. Hebeler, J. D. Holt, G. R. Jansen, M. Kowalska, K. Kreim, W. Nazarewicz, R. Neugart,
G. Neyens, W. N\"{o}rtersh\"{a}user, T. Papenbrock, J. Papuga
, A. Schwenk, J. Simonis, K. A. Wendt and D. T. Yordanov, Nat. Phys. {\bf 12}, 594 (2016).
\bibitem{Gorges} C.Gorges, L. V. Rodr$\acute{\mbox{i}}$guez, D. L. Balabanski, M. L. Bissell, K. Blaum, B. Cheal, R. F. Garcia Ruiz, 
G. Georgiev, W. Gins, H. Heylen, A. Kanellakopoulos, S. Kaufmann, M. Kowalska, V. Lagaki, S. Lechner, 
B. Maa{\ss}, S. Malbrunot-Ettenauer, W. Nazarewicz, R. Neugart, G. Neyens, W. N\"{o}rtersh\"{a}user, P.-G. Reinhard, 
S. Sailer, R. S$\acute{\mbox{a}}$nchez, S. Schmidt, L. Wehner, C. Wraith, L. Xie, Z. Y. Xu, X. F. Yang, and D. T. Yordanov. Phys. Rev. Lett. {\bf 122}, 192502 (2019).
\bibitem{Miller} A. J. Miller, K. Minamisono, A. Klose, D. Garand, C. Kujawa, J. D. Lantis, Y. Liu, B. Maa{\ss},
P. F. Mantica, W. Nazarewicz, W. N\"ortersh\"auser, S. V. Pineda, P.-G. Reinhard, D. M. Rossi,
F. Sommer, C. Sumithrarachchi, A. Teigelhöfer and J. Watkins, Nat. Phys. 15, 432(2019)
\bibitem{ma2016} X.F. Li, D.Q. Fang and Y.G. Ma, Nucl. Sci. Tech. {\bf 27}, 71 (2016).
\bibitem{5a} I. Angeli and K. P. Marinova, At. Data Nucl. Data Tables 99, 69
(2013).
\bibitem{6a}N. J. Stone, J. Phys. Chem. Ref. Data {\bf 44}, 031215 (2015).
\bibitem{caurier}E. Caurier, K. Langanke, G. Martinez-Pinedo, F. Nowacki and P. Vogel, Phys. Lett. B {\bf 522}, 240 (2001).
\bibitem{Vermeeren}L. Vermeeren, R. E. Silverans, P. Lievens, A. Klein, R.
Neugart, C. Schulz, and F. Buchinger, Phys. Rev. Lett. {\bf 68}, 1679 (1992).
\bibitem{Anselment}M. Anselment, W. Faubel, S. G\"{o}ring, A. Hanser, G. Meisel,
H. Rebel, and G. Schatz, Nucl. Phys. {\bf A451}, 471 (1986).
\bibitem{Cocolios}T. E. Cocolios, W. Dexters, M. D. Seliverstov, A. N. Andreyev, S. Antalic, A. E. Barzakh, B. Bastin, J. B\"{u}scher, I. G. Darby, D. V. Fedorov, V. N. Fedosseyev, K. T. Flanagan, S. Franchoo, S. Fritzsche, G. Huber, M. Huyse, M. Keupers, U. K\"{o}ster, Yu. Kudryavtsev, E. Man$\acute{\mbox{e}}$, B. A. Marsh, P. L. Molkanov, R. D. Page, A. M. Sjoedin, I. Stefan, J. Van de Walle, P. Van Duppen, M. Venhart, S. G. Zemlyanoy, M. Bender, and P.-H. Heenen, Phys. Rev. Lett. {\bf 106}, 052503 (2011).
\bibitem{Forssen} C. Forss$\acute{\mbox{e}}$n, E. Caurier, P. Navr$\acute{\mbox{a}}$til, Phys. Rev. C {\bf 79}, 021303(R)(2009).
\bibitem{stoitsov}M. V. Stoitsov, J. Dobaczewski, W. Nazarewicz, S. Pittel, and D.J. Dean, Phys. Rev. C {\bf 68}, 054312 (2003).
\bibitem{goriely}S. Goriely, N. Chamel, and J. M. Pearson, Phys. Rev. C {\bf 82}, 035804 (2010).
\bibitem{nakada06}H. Nakada, Nucl. Phys. A {\bf 764}, 117 (2006).
\bibitem{Reinhard} P.-G. Reinhard, W. Nazarewicz, Phys. Rev. C {\bf 95}, 064328(2017).
\bibitem{nakada19} H. Nakada, Phys. Rev. C {\bf 100}, 044310(2019).
\bibitem{lalazissis}G. A. Lalazissis, S. Raman, and P. Ring, At. Data Nucl. Data Tables {\bf 71}, 1 (1999).
\bibitem{zhao}P. W. Zhao, Z. P. Li, J. M. Yao, and J. Meng, Phys. Rev. C {\bf 82}, 054319 (2010).
\bibitem{duflo}J. Duflo, Nucl. Phys. A {\bf 576}, 29 (1994).
\bibitem{GK}J. Piekarewicz, M. Centelles, X. Roca-Maza, and X. Vi$\tilde{\mbox{n}}$as, Eur. Phys. J. A {\bf 46}, 379 (2010)
\bibitem{zhang}S. Zhang, J. Meng, S.-G. Zhou, and J. Zeng, Eur. Phys. J. A {\bf 13}, 285 (2002).
\bibitem{wangn}N. Wang and T. Li, Phys. Rev. C {\bf 88}, 011301(R) (2013).
\bibitem{sheng}Z. Sheng, G. Fan, J. Qian, and J. Hu, Eur. Phys. J. A {\bf 51}, 40 (2015).
\bibitem{bao}M. Bao, Y. Lu, Y. M. Zhao, and A. Arima, Phys. Rev. C {\bf 94}, 064315 (2016).
\bibitem{sun}B. H. Sun, Y. Lu, J. P. Peng, C. Y. Liu, and Y. M. Zhao, Phys. Rev. C {\bf 90}, 054318 (2014).
\bibitem{Utama2016} R. Utama, J. Piekarewicz, and H.B. Prosper, Phys. Rev. C {\bf 93}, 014311(2016).
\bibitem{14} Z.M. Niu and H.Z. Liang, Phys. Lett. B {\bf 778} (2018) 48.
\bibitem{Wu} X.H. Wu, P.W. Zhao, Phys. Rev. C {\bf 101}, 051301(R)(2020).
\bibitem{ma}Y.F. Ma, C. Su, J. Liu, Z.Z. Ren, C. Xu and Y.H. Gao, Phys. Rev. C {\bf 101}, 014304 (2020).
\bibitem{utama}R. Utama, W.-C. Chen, and J. Piekarewicz, J. Phys. G {\bf 43}, 114002 (2016).
\bibitem{Neufcourt2018} L. Neufcourt, Y. Cao, W. Nazarewicz, and F. Viens, Phys. Rev. C {\bf 98}, 034318(2018).
\bibitem{Neufcourt2019} L. Neufcourt, Y. Cao, W. Nazarewicz, E. Olsen, and F. Viens, Phys. Rev. Lett. {\bf 122}, 062502(2019).
\bibitem{15} Z.M. Niu, H.Z. Liang, B.H. Sun, W.H. Long and Y.F. Niu, Phys. Rev. C {\bf 99}, 064307 (2019).
\bibitem{16} Zi-Ao Wang, Junchen Pei, Yue Liu and Yu Qiang, Phys. Rev. Lett. {\bf 123}, 122501 (2019).
\bibitem{ma2020} C.W. Ma, D. Peng, H.L. Wei, Z.M. Niu, Y.T. Wang and R. Wada, Chin. Phys. C {\bf 44}, 014104 (2020).
\bibitem{Lasseri} R.D. Lasseri, D. Regnier, J.P. Ebran, and A. Penon, Phys. Rev. Lett. {\bf 124}, 162502(2020).
\bibitem{18} T. Tieleman and G. Hinton. COURSERA: Neural networks for machine learning, {\bf 4(2)} 26 (2012).
\bibitem{angeli2015} I. Angeli and K. Marinova, J. Phys.: Conf. Ser. {\bf 724}, 012032 (2016).
\bibitem{cakirli} R. B. Cakirli, R. F. Casten, and K. Blaum, Phys. Rev. C {\bf 82}, 061306(R) (2010).
\bibitem{raman} S. Raman, C. W. Nestor Jr. and K. H. Bhatt, Phys. Rev. C {\bf 37}, 805 (1988).
\bibitem{Weizs} C. F. V. Weizsacker, Z Phys. {\bf 96}, 431 (1935).
\bibitem{bethe} H. A. Bethe, R. F. Bacher, Rev. Mod. Phys. {\bf 8}, 82 (1936).
\bibitem{fricke} G. Fricke and K. Heilig, $Nuclear~Charge~Radii$, Group I: Elementary Particles, Nuclei and Atoms Vol. 20 (Springer, New York, 2004).
\bibitem{hfbrad} K. Bennaceur, J. Dobaczewski, Comput. Phys. Commun. {\bf 168}, 96 (2005).
\bibitem{inakura} T. Inakura and H. Nakada, Phys. Rev. C {\bf 92}, 064302 (2015).
\bibitem{colo} G. Col$\grave{\mbox{o}}$ and X. Roca-Maza, Acta Phys.Polon. B {\bf 46}, 395 (2015).
\bibitem{li2008} B.A. Li, L.W. Chen and C.M. Ko, Phys. Rep. {\bf 464}, 113 (2008).
\bibitem{yu2020} H. Yu, D.Q. Fang and Y.G. Ma, Nucl. Sci. Tech. {\bf 31}, 61 (2020).
\bibitem{roca} X. Roca-Maza, M. Brenna, B. K. Agrawal, P. F. Bortignon, G. Col$\grave{\mbox{o}}$, Li-Gang Cao, N. Paar, and D. Vretenar, Phys. Rev. C {\bf 87}, 034301 (2013).
\bibitem{roca2012} X. Roca-Maza, G. Col$\grave{\mbox{o}}$, and H. Sagawa, Phys. Rev. C {\bf 86}, 031306(R) (2012).
\bibitem{roca1} X. Roca-Maza, private communications.
\end{thebibliography}
\end{document}